\documentclass[doublecol,sd,amsmath,amssymb,amstext]{epl2}
\usepackage{color}
\usepackage{graphicx}
\usepackage{graphicx,epstopdf}
\usepackage{dcolumn}
\usepackage{bm}
\usepackage{mathrsfs}
\usepackage{amsfonts}
\usepackage{amsmath}

\begin{document}

\title{Perspective on attractive-repulsive interactions in dynamical networks: progress and future }
\author{Soumen Majhi, Sayantan Nag Chowdhury, and Dibakar Ghosh \footnote{Email: diba.ghosh@gmail.com}}
\shortauthor{S. Majhi, S. Nag Chowdhury, and D. Ghosh }
\institute{Physics and Applied Mathematics Unit, Indian Statistical Institute, 203 B. T. Road, Kolkata-700108, India}

\date{\today}

\abstract{Emerging collective behavior in complex dynamical networks depends on both coupling function and underlying coupling topology. Through this perspective, we provide a brief yet profound excerpt of recent research efforts that explore how the synergy of attractive and repulsive interactions influence the destiny of ensembles of interacting dynamical systems. We review the incarnation of collective states ranging from chimera or solitary states to extreme events and oscillation quenching arising as a result of different network arrangements. Though the existing literature demonstrates that many of the crucial developments have been made, nonetheless, we come up with significant routes of further research in this field of study. }

 {Complex systems }

\maketitle



\par Emergence of diverse macroscopic states in ensembles of interacting oscillators depending on coupling configuration is a central issue of interest in many different
fields of research. Attractive (positive) coupling, in general, gives rise to in-phase allignment among the oscillators. On the other hand, repulsive (negative) coupling drives the oscillators apart and induces out-of-phase synchronization. However, realistic systems are far more complicated and introducing mixed coupling with both positive and negative couplings is another way of bringing the real coupled-oscillator systems closer to reality. Simultaneity of attractive and repulsive couplings can be observed in a plethora of different contexts including sociology \cite{martins2009divide}, ecology \cite{giron2016synchronization,bacelar2014exploring} as well as in modelling of physical \cite{sun2016realization,dixit2020static}, biological \cite{daido1987population}, socio-technical systems \cite{burylko2012competition,el2013synchronization,el2013phase}, and thus it serves as a simple yet highly efficient framework to understand the underlying mechanism of many complex systems. Coexistence of these couplings makes the system frustrated \cite{zanette2005synchronization} and examples of such frustrated systems are omnipresent. Even the most complex organ brain consists of attractive and repulsive couplings \cite{myung2015gaba,izhikevich2007dynamical} and it contains almost $75\%$ excitable neurons and $25\%$ inhibitory neurons \cite{soriano2008development, vogels2009gating}.

\begin{figure}[!t]
	\centerline{\includegraphics[scale=0.3100]{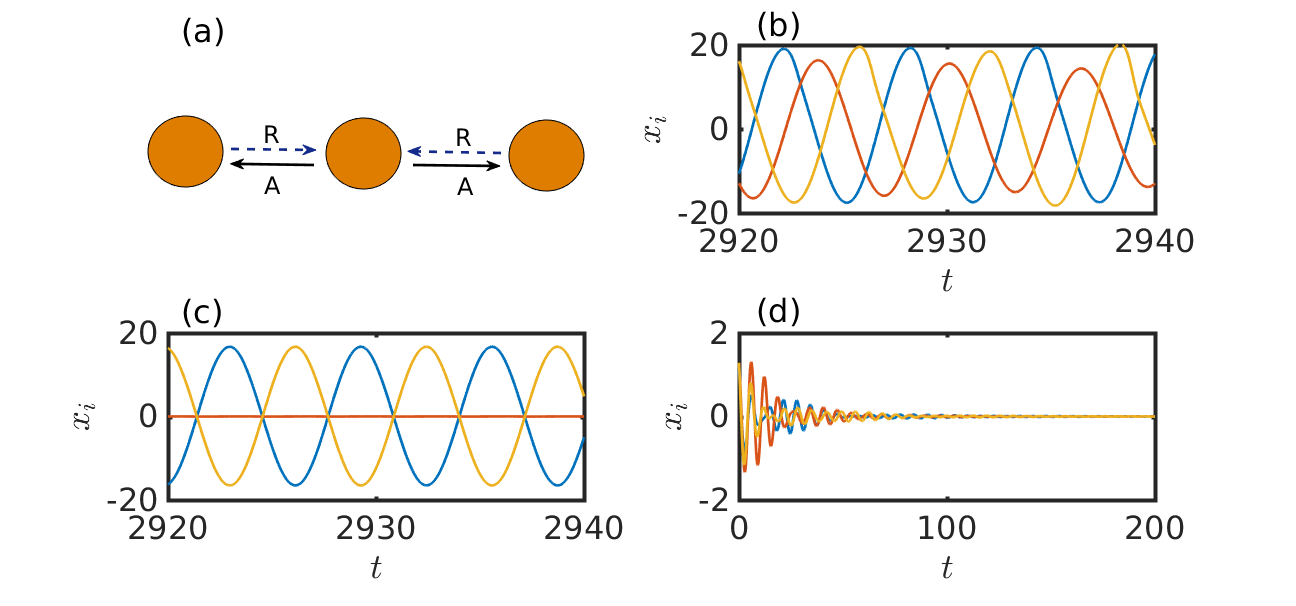}}
	\caption{The transition mechanism for the emergence of AD with increasing coupling strength $\epsilon$ for a chain of $N=3$ oscillators is explored. (a) A schematic presentation is portrayed for the relay system, where `R' (dashed lines) stands for repulsive sum feedback  and `A' (solid lines) signifies attractive diffusive interaction. (b) Incoherent behaviors of $x_1$, $x_2$ and $x_3$ are depicted through oscillations with different amplitudes for $\epsilon=0.04$. (c) Outer-most oscillators ($x_1$ and $x_3$) are in anti-synchronized state, but the amplitude of the middle oscillator ($x_2$) is damped out to the AD state whenever $\epsilon=0.09$. (d) All the oscillators arrive at the AD state at $\epsilon=0.14$. All the panels are drawn with fixed initial condition $(x_1(0),y_1(0),z_1(0),x_2(0),y_2(0),z_2(0),x_3(0),y_3(0),z_3(0))=(1.1,0.2,0.3,1.2,0.3,0.4,1.3,0.5,0.6)$. We refer the reader to the ref.\ \cite{zhao2018amplitude} for further details. }\label{fig3}
	\end{figure}

\par Interplay between attractive and repulsive couplings may originate suppresion (death) of oscillation among coupled oscillators. Along this line, the number of minimal repulsive links, which is sufficient enough to induce death in a network of globally and diffusively coupled Stuart-Landau (SL) oscillators, is inspected in the ref.\ \cite{hens2013oscillation}. Their numerical investigation attests that the repulsive strength should be passed through at least $30\%$ links of the network for a death scenario to emerge. Zhao et al. \cite{zhao2018amplitude} found that while uniform coupling (i.e., either only attractive or only repulsive alone) is unable to stabilize the amplitude death (AD) state, mixed coupling can induce AD in the relay system of SL and R\"{o}ssler oscillators. Figure \ref{fig3} demonstrates the tranisition mechanism for the occurence of AD in a relay system of three R\"{o}ssler oscillators, where the middle oscillator is repulsively coupled with the outer oscillators and the outer oscillators are attractively coupled with the middle oscillator. The dynamical equation of the $i$-th R\"{o}ssler oscillator ($i=1,2,3$) is given by

\begin{equation}\label{2}
\dot{x_i}=-y_i-z_i+\epsilon G_i,\\
\dot{y_i}=x_i+0.1y_i,\\
\dot{z_i}=0.1+z_i(x_i-14).
\end{equation}

Here, $\epsilon$ is the coupling strength and $G_2=[(x_1-x_2)+(x_3-x_2)]$ is the difference feedback between two neighbouring oscillators. The repulsive sum feedback is depicted through $G_1=-(x_2+x_1)$ and $G_3=-(x_2+x_3)$. The system settles down to AD through three basic steps. Initially with small coupling strength $\epsilon=0.04$, the three oscillators are oscillating with different amplitudes (fig.\ \ref{fig3}(b)). The amplitude of the middle oscillator is smaller compared to the outer oscillators. For $\epsilon > \epsilon_c=0.1$, suppression of oscillations to AD state is observed as in fig.\ \ref{fig3}(d). For an intermediate choice of $\epsilon$, the outer oscillators maintain anti-synchronization while the middle oscillator acheives AD (fig.\ \ref{fig3}(c)). The observed transition mechanism for the emergence of AD remains unchanged if the relay system is coupled in attractive-repulsive-attractive (ARA) manner instead of RAR way as shown in fig.\ \ref{fig3}. Emergence of amplitude death in a network of identical oscillators through repulsive mean coupling under coupling delay is reported in the ref.\ \cite{bera2016emergence}. A globally coupled network of SL and R\"{o}ssler oscillators under the effect of time delay can reveal oscillation quenching in the form of AD or oscillation death (OD), whenever suitable oscillators are perturbed through repulsive diffusive delay coupling \cite{kundu2019emergent}. Revival of oscillation is also possible from the AD state depending on the internal parameters of the network. The transition mechanism from AD to OD under attractive coupling and with additional repulsive link is investigated in the ref. \cite{hens2014diverse} by considering three different systems containing two or three identical oscillators. The detailed analysis of synchronized and antisynchronized oscillatory states along with the OD state has been presented, and these collective behaviors emerge as a result of the trade-off between attractive and repulsive couplings \cite{sathiyadevi2017spontaneous}. 
\par Earlier, the fascinating outcomes that emerge due to attractive-repulsive interactions are also revealed by Hong et al.\ \cite{hong2011kuramoto} through their studies. They consider a generalization of the Kuramoto model,

\begin{equation} \label{1}
\dot{\phi_i}=\omega_i+\frac{K_i}{N} \sum\limits_{k=1}^{N}\sin({\phi_k-\phi_i}),~~~i=1,2,...,N,
\end{equation}
with $Re^{j\Phi}=\frac{1}{N}\sum\limits_{k=1}^{N}e^{j\phi_k}$, which reduces to
\begin{equation} \label{1q}
\dot{\phi_i}=\omega_i+K_i R \sin({\Phi-\phi_i}),~~~i=1,2,...,N,
\end{equation}
where the natural frequencies $\omega_i$ are drawn from a Lorentzian probability density $g(\omega)=\gamma/[\pi(\omega^2+\gamma^2)]$ of width $\gamma$ and mean $\langle \omega \rangle =0$. The mean-field variables $R$ and $\Phi$ describe the phase coherence and average phase, respectively. The interaction strength among those oscillators  $K_i$ is drawn from a double delta distribution $\Gamma(K)=p \delta(K-D_2)+(1-p)\delta(K-D_1)$, where $D_2 > 0$ and $D_1<0$. Motivated by sociophysical models of opinion formation, the repulsively coupled oscillators represent contrarians who oppose everything, while the positively coupled oscillators are treated as conformist favoring coherence in the population. The probability of being a conformist is $p$. Clearly, the limiting cases reflect to the scenario that either the system is full of contrarians for $p=0$ or the system coincides with the original Kuramoto model for $p=1$. When $\gamma$ is sufficiently small, the system converges to an asymptotic behavior termed as \textit{traveling wave state} for an intermediate regime of $p$. The phase distribution at this state follows a constant distance $d \ne \pi$ possessing a non-zero mean-phase velocity $\langle \dot{\phi_i} \rangle \ne 0$. For small $p$, the system is dominated by contrarians and as a result of that, the globally coupled phase oscillators are completely desynchronized. As soon as $p$ crosses a certain threshold, the system settles down to a \textit{$\pi$-state}, where both conformist and contrarian exhibit stationary distribution of phases leading to a fixed point behavior of the order parameter $Re^{j\Phi}$. The peaks of both distributions maintain a constant mean phase difference $d=\pi$. Further increment of $p$ generates the traveling wave state before the system collapses back to the partially synchronized $\pi$-state. Note that the coupling strengths in Eq.\ \eqref{1} appear outside of the summation. Whenever the coupling strengths are inserted within the summation, the traveling wave state and $\pi$-state are no longer observed \cite{hong2012mean}. Unexpectedly, the system exhibits a second-order phase transition similar to the Kuramoto oscillator. A different situation \cite{hong2016correlated} can be analyzed when the correlation between the natural frequencies and coupling strengths is established deterministically by distributing an equal number of positively and negatively coupled oscillators around $\omega=0$. This correlated disorder ultimately favors the partially locked state for any non-zero fraction of positively coupled oscillators. To understand the local dynamics around each fixed point, the eigenvalues of the Jacobi matrix around each fixed point in the Hong-Strogatz model are explicitly calculated in the ref.\ \cite{peng2015analysis}.

\par Motivated by the Daido's pioneering work \cite{daido1992quasientrainment}, where a new type of ordered state analogous to glass transition in a large ensemble of coupled limit-cycle oscillators with positive and negative couplings is explored, Hong et al.\ \cite{hong2011conformists} investigated a variant of Kuramoto model with assymetric pairwise interaction and uniform natural frequency. This asymmetric interaction creates different types of frustration as the $l$-th oscillator may be negatively coupled with the $j$-th oscillator, but in return the $j$-th one is coupled positively to the $l$-th one. The percieved numerical and analytical techniques reveal that the long-time dynamics for the homogeneous system converges to one of the following four states: (i) incoherent state, (ii) \textit{blurred $\pi$-state}, (iii) traveling wave state, and (iv) $\pi$-state. The same model of identical phase oscillators is analyzed with a phase shift and arbitrary finite number of oscillators causing rich complex dynamical behavior \cite{burylko2014bifurcation}. The presence of weak pinning force in the model \cite{hong2011conformists} helps to produce several peculiar dynamical states including \textit{periodic synchronization}, \textit{breathing chimera} and \textit{fully pinned state} depending on the fraction of the conformists \cite{hong2014periodic}. If the pinning force is strong enough, then only the fully pinned state exists in the system. The collective behavior of the generalized Kuramoto model with an external pinning force \cite{yuan2018periodic} is also investigated under the limelight of the situation, in which the natural frequencies of the oscillators follow a uniform probability density. Diverse emergent behavior including traveling wave state, $\pi$-state, blurred $\pi$-state and periodic synchronous state (termed as \textit{oscillating $\pi$-state}) can be obtained due to the interplay of conformists and contrarians. Yuan et al. \cite{yuan2016dynamics} also found such rich dynamics in a variant of the generalized Kuramoto model with a bi-harmonic coupling function term, in which oscillators with positive first harmonic coupling strength are conformists and oscillators with negative first harmonic coupling strength are contrarians. Depending on the parameters, Kuramoto model of globally coupled phase oscillators with time-delayed positive and negative couplings is also capable of displaying a variety of dynamic behaviors including fully coherent, incoherent states and mixed (coherent, incoherent, and clustered) states \cite{wu2018dynamics}. The recent advances on the emergence of a traveling state has been studied with positive and negative couplings \cite{choi2014dynamics,choi2015traveling,choi2019effects} and thus offering new possibilities for exploration. \nolinebreak
\par In fact, coherent motion is not necessarily the desired state always, e.g., wobbling of the millennium bridge and traffic congestions in networks. A fraction of contrarians is significant enough to suppress the global synchronization of the system. Based on the local information, the coherent behavior can be reduced effectively whenever the number of contrarians crosses a certain threshold \cite{louzada2012suppress}. Surprisingly, global information may still entertain the global synchronization state as illustrated in the ref.\ \cite{louzada2012suppress}. Zanette \cite{zanette2005synchronization} inspected the frustration in a model with pairwise coupling analogous to the magnetic XY model in the limit of $\omega_i=0$. From a different perspective in the ref.\ \cite{bonilla1993glassy}, appearance of the stable \textit{glassy phase} state and the \textit{mixed} state in a model of phase-coupled frustrated oscillators with random excitatory and inhibitory couplings of Van Hemmen type along with incoherent and synchronized phases is contemplated in the thermodynamic limit. The dynamical robustness property of the damaged networks under the influences of both repulsive and attractive couplings has been recently inquired in the ref.\ \cite{bera2020additional}. Also, the presence of a suitable number of repulsive links in a system of globally coupled Van der Pol oscillators diminishes the coherent behavior and leads to an enhanced response of the external signal \cite{vaz2011synchronisation}.

\par
\begin{figure}[!t]
	\centerline{\includegraphics[scale=0.3100]{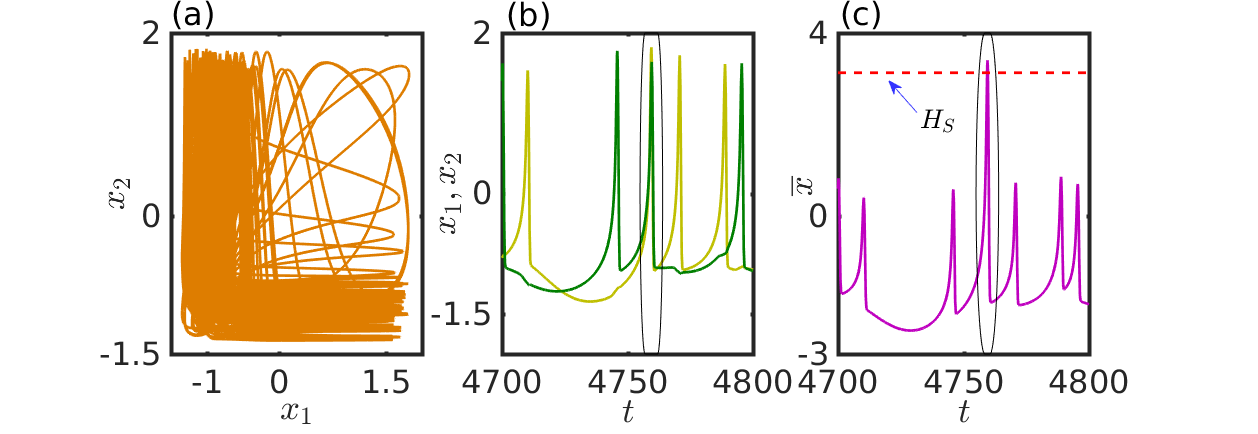}}
	\caption{(a) Plot of $x_1$ vs. $x_2$ illustrates the fact that they are in out of phase oscillations for most of the time. Occasionally, they traverse towards the in-phase synchronization manifold. (b)-(c) The temporal behavior of $x_1$ and $x_2$ reflects that when one oscillator displays bursting oscillations, then the other one is in the quiescent state. However, a scenario is highlighted by an elliptical mark, where spikings of these two oscillators coincide. This intermittent overlapping ultimately reveals a high-amplitude oscillation of the observable $\overline{x}=x_1+x_2$ in the temporal domain. The dashed line in panel (c) is $H_S \approx 3.1333$. All the figures are simulated with $k_1=0.07$ and $k_2=-0.08$. The initial condition $(x_1(0),y_1(0),z_1(0),x_2(0),y_2(0),z_2(0))=(0.1,0.2,0.3,0.6,0.7,0.8)$ is also kept fixed. For further details, please see the ref.\ \cite{mishra2018dragon}. }\label{fig2}
\end{figure}

 Interestingly, strong coherence can still be observed in neuronal networks even with the presence of both excitable and inhibitory neurons \cite{borgers2003synchronization}. A tit-for-tat strategy is implemented to disregard the negative role of contrarian oscillators and to increase the synchronization in the ref.\ \cite{zhang2013efficient}. A small fraction of phase-repulsive links can help to sustain and enhance synchronization in a small-world network composed of nonidentical coupled units \cite{leyva2006sparse}. Restrepo et al.\ \cite{restrepo2006synchronization} unveiled a theoretical approximation to find the critical coupling strength, where a macroscopic transition to synchronization takes place in a large directed network of phase oscillators with mixed positive-negative coupling. Scale-free neuronal networks with attractive or phase-repulsive coupling and finite delay lengths have been studied in the ref.\ \cite{wang2011synchronous}. In the ref.\ \cite{mishra2018dragon}, two periodically bursting Hindmarsh-Rose (HR) neurons are considered given by the following equations 
\begin{equation}\label{4}
\begin{split}
\dot{x}_j=y_j+3x_j^2-x_j^3-z_j+4-k_j(x_j-2)\Gamma(x_l),\\
\dot{y}_j=1-5x_j^2-y_j,\hspace{4.3cm}\\
\dot{z}_j=0.01[5(x_j+1.6)-z_j]; \hspace{0.3cm} j,l=1,2~(j \ne l).
\end{split}
\end{equation} 
Here, the chemical synaptic interaction is captured by the sigmoidal function $\Gamma(x_l)={\big[1+\exp[-10(x_l+0.25)]\big]}^{-1}$. The coupling strengths $k_j$ decide the type of interaction. Here, we choose $k_1=0.07$ and $k_2=-0.08$. These combination of excitatory and inhibitory interactions is capable of generating \textit{extreme events} as shown in fig.\ \ref{fig2}. The phase portrait given in fig.\ \ref{fig2}(a) draws the conclusion that the variables endure out-of-phase oscillation for most of the time, and occasionally they travel towards the in-phase synchronization $x_1=x_2$ manifold. Coincidence of two such spikes are highlighted in fig.\ \ref{fig2}(b) through an elliptical mark. This intermittent unison generates a large amplitude oscillation in the temporal dynamics of $\overline{x}=x_1+x_2$ (cf. fig.\ \ref{fig2}(c)). The probability distribution function of $\overline{x}$ is found to display Dragon-king distribution revealing the appearence of Dragon-king-like extreme events. The dashed line in fig.\ \ref{fig2}(c) is the extreme event indicating threshold $H_S=\mu+6\sigma$, where $\mu$ and $\sigma$ are the mean value and the standard deviation of all the peak values in a sufficiently long time series of $\overline{x}$, respectively.

\par But, the agents of the social systems rarely remain isolated and the strategies of those agents generally change over time to avoid the undesired phase-locked state. Motivated by these facts, fresh ideas emerge among researchers and few time-varying networks \cite{chowdhury2019extreme,chowdhury2020distance} are contemplated with attractive-repulsive interactions, which manifest extreme events for an intermediate choice of coupling strengths. During the transition from synchronization to incoherent regime, those systems of mobile agents can give rise to extreme events through the route of on-off intermittency. Moreover, such a competing interaction due to the interplay of positive inter-layer and negative intra-layer interactions may initiate \textit{solitary states} in multiplex networks of coupled oscillators \cite{majhi2019solitary}. In this state, one or a few units of the ensemble split off and behave differently from the other units. Specifically, through this paper, Majhi et al.\ \cite{majhi2019solitary} articulated the emergence of such a weak chimera-like dynamical state in a bi-layer multiplex network exhibiting competitive interactions in terms of the opposite characteristics of inter- and intra-layer couplings. Diverse patterns of solitary states with cluster synchronization and oscillation death states are encountered dealing with the FitzHugh-Nagumo system in its equilibrium and periodic regimes. For the equilibrium regime, the FitzHugh-Nagumo systems on the two layers are assumed on the two sides of the Hopf bifurcation in response to the external stimulus, along with positive inter-layer and negative intra-layer strengths. Evidence of such a peculiar phenomenon has been presented while contemplating with the Lorenz system in its periodic and chaotic regimes.

\begin{figure*}[!htb]
	\centerline{\includegraphics[scale=0.3200]{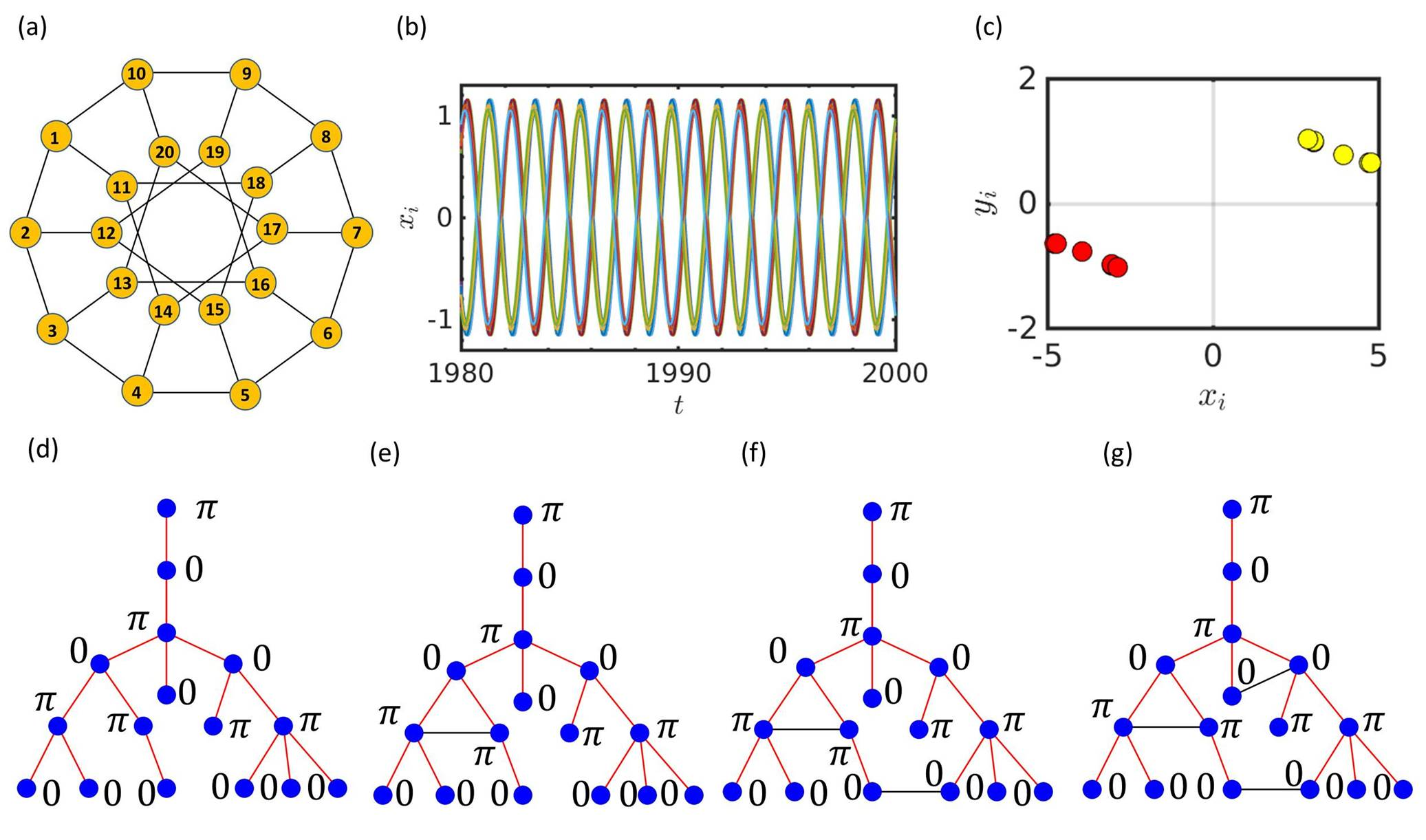}}
	\caption{(a)-(c){\bf Anti-phase synchronization in bipartite graph}: (a) The Desargues graph with $N=20$ vertices and $L=30$ edges is considered to demonstrate the manifestation of anti-phase synchronization in a bipartite graph. The vertices of this balanced bipartite graph can be partitioned into $U=\{1,3,5,7,9,12,14,16,18,20\}$ and $V=\{2,4,6,8,10,11,13,15,17,19\}$ respectively. (b) A spanning tree of the connected network is accounted for and the repulsive coupling is spread through that spanning tree with $k_R=-0.1$. The complementary subgraph with $11$ links is positively coupled with coupling strength $k_A=0.001$. The system settles down to anti-phase synchronization while the oscillators within the set $U$ and $V$ are in-phase synchronized between themselves. (c) Higher negative coupling with $k_R=-4.0$ helps to reach the entire network to different inhomogeneous steady states with $F=0$ under suitable initial conditions. (d)-(g) {\bf Construction of a network with pre-specified $F$}: A connected acyclic undirected graph with $N=16$ vertices and $L=15$ edges is given in the panel (d). Initially, all links are repulsively coupled (red lines). By joining one by one attractive edges (black lines), a new non-bipartite network is designed with $F_{desired}=\frac{1}{3}$ in the panel (g). We refer the reader to the ref.\ \cite{chowdhury2020effect} where further details of the simulations can also be found. }\label{fig1}
\end{figure*}
\par A transition from two-cluster synchronization to partial synchronization in a globally coupled phase oscillators can be realized due to the interplay between attractive and repulsive interactions within two groups of identical oscillators \cite{teichmann2019solitary}, where the groups differ in their natural frequencies. If this frequency mismatch between attractive and repulsive units is smaller than some critical value, then the system may exhibit solitary states. Jalan et al.\ \cite{jalan2019inhibition} found that an inhibitory layer of negatively coupled nodes hinders the formation of synchronized giant cluster in the excitatory layer of all positively coupled Kuramoto oscillators resulting in the manifestation of \textit{explosive synchronization} (ES) in the multiplex network.  An efficient approach in order to convert the first order transition to a second order transition is proposed in the ref.\ \cite{zhang2016suppressing}. By changing a small fraction of oscillators into the contrarians depending on the average degree and the network size, one can easily suppress ES in a network of coupled Kuramoto oscillators. The dynamics of a model with interlinked
positive and negative feedback loops is explored in the ref.\ \cite{tian2009interlinking}. Several rich complex phenomenon including monostability, bistability, excitability and oscillation can be observed in this system (which can be thought of as a tunable motif) by changing the feedback strength. The heterogeneity in the coupling and in the characteristic timescales may influence the system behavior and may generate a situation which is impossible without timescale differences. Kirillov et al.\ \cite{kirillov2020role} inspected a heterogeneous ensemble of two groups with different internal timescales. One of these group possesses attractive coupling and the other one is repulsively coupled. Although qualitatively same behavior with the equal timescales is observed for the slower attractive group, in contrast when the attractive group is faster than the repulsive group, then the emergence of new dynamical regimes including bistability and rotation of the group mean field with respect to each other is found. In fact one of the recent findings suggest that instead of uniform couplings (purely positive or purely negative couplings), mixed positive-negative coupling may be helpful for the signal amplification \cite{liang2020positive}. With appropriate choices of the ratio between two types of coupling and the coupling strength, the system of globally coupled bistable oscillators subject to a common weak external signal can lead to resonance like behavior and the dynamics of the system settles down to the three oscillation clusters.

\par The emergence of bipolar aggregations for the two sub-ensembles
of the swarm sphere model under attractive-repulsive couplings has been explored recently \cite{ha2020emergence}. The coherence-incoherence transition in networks of globally coupled identical oscillators with attractive and repulsive interactions is found to occur through the appearence of solitary states, provided the attractive and repulsive groups act in antiphase or close to that \cite{maistrenko2014solitary}. Recently, Chowdhury et al.\ \cite{chowdhury2020effect} have made an effort to figure out whether there exist any universal generic path in a connected network of attractively and diffusively coupled SL oscillators, which will help to attain \textit{anti-phase synchronization} by passing a decent amount of repulsive strengths through it. Using the $0$-$\pi$ rule and bifurcation analysis, they showed that the anti-phase synchrnization is possible in any connected network if and only if the network is bipartite in nature. A measure $F=\langle \frac{1}{L} \sum_{i<j} A_{ij}[1+\cos(\theta_i-\theta_j)] \rangle$ is used to determine whether each pair of adjacent nodes follows antiphase states or not. Here, $A_{ij}$ is the $N \times N$ adjacency matrix of the network and $L$ is the total number of links. $\theta_i$ is the intrinsic phase of the $i$-th oscillator. Clearly, $F$ lies within $[0,2]$, and particularly $F=0$ reflects the emergence of antiphase synchronization $($i.e.\ $|\theta_i-\theta_j|=\pi)$. This measure $F$ thus acts like a unique fingerprint which will able to distinguish between bipartite $(F=0)$ and non-bipartite graphs $(F>0)$. If an adequate amount of repulsive strengths can be passed through any of the spanning trees of a connected bipartite network, then the system may split into two clusters maintaining a phase difference of $\pi$. To demonstrate this feature, the only known non-planar partial cube Desargues graph (see fig.\ \ref{fig1}(a)) is considered. The connectedness of this graph assures the existence of a spanning tree of this network and the bipartiteness of this graph ultimately favors anti-phase synchronization, whenever the negative coupling is passed through any of the existing spanning trees embedded in the considered graph. The positively-negatively coupled limit cycle oscillators under this arrangement exhibit a scenario where the phase differences between the existing links are in the difference of $\pi$. Although, the system acquires zero frustration $(F = 0)$ with oscillation states (fig.\ \ref{fig1}(b)) and fixed points (fig.\ \ref{fig1}(c)), respectively, however the system becomes multistable and hence careful selection of initial conditions is needed in order to achieve anti-phase synchronization. In fact, the basin of attraction for $F=0$ becomes narrower with increasing network size. This understanding is also recognized in another recent study \cite{vathakkattil2020limits} where the anti-phase synchronization is found to be limited to small-sized networks due to its dependencies on several factors including connectivity of the network, strength of interaction over distance, and symmetry of the network. Construction of a non-bipartite graph by adding few attractive links from a given repulsive tree with desired $F$ is exemplified in figs.\ \ref{fig1}(d)-(g) using the algorithm prescribed in the ref.\ \cite{chowdhury2020effect}. Initially, a repuslive tree of $16$ nodes with coupling strength $k_R=-4.0$ is given in fig.\ \ref{fig1}(d). From this theory, one can create a sparse graph from this non-frustrated graph $(F=0)$ with $F_{desired}=\frac{1}{3}$. Then, one only needs to decompose the bipartite graph into two disjoint sets, say $U$ and $V$ and add a link between two nodes either from the set $U$ or from $V$, so that there is no link in between those nodes at prior. Using the proposed $0$-$\pi$ rule, $F=\frac{2m}{L}$ is calculated, where $m=L-(N-1)$ is the number of attractive links. If $F<F_{desired}$, then one needs to add again a link between two nodes either from the set $U$ or from $V$. This process will continue until the desired $F$ is achieved. By adding only $3$ attractive links for the graph given in fig.\ \ref{fig1}(d), one can accomplish their motive as shown in fig.\ \ref{fig1}(g).

\par On the other hand, two different types of chimera-like behavior have been detected in a network of globally coupled Li\'{e}nard system under attractive and repulsive mean-field feedbacks \cite{mishra2015chimeralike}. Diverse collective states in the form of cluster chimera death and solitary state are observed in nonlocally coupled oscillatory systems with attractive and repulsive couplings \cite{sathiyadevi2018distinct}. Stable amplitude chimera and traveling wave states are encountered in nonlocally coupled network of oscillators in presence of both attractive and repulsive interactions \cite{sathiyadevi2018stable}. In the following, we discuss about a significant observation of frequency-modulated chimera-like pattern during explosive transitions to synchronization in networks of the heterogeneous Kuramoto model \cite{frolov2020chimera}. Interestingly, this chimera-like behavior has been encountered \textit{not} for any induced repulsion in the networked system, rather it has been shown that this chimera-like behavior emerges due to the coexistence of \textit{evolved} attractively and repulsively coupled subpopulations of oscillators. A network of Kuramoto phase oscillators is considered as follows:
\begin{equation}
	\begin{split}
		\dot{\phi}_i & = \omega_i + \lambda R_i \sum_{l=1}^{N} A_{il} \sin (\phi_l - \phi_i ),~~~i=1,2,...,N,\\
		R_i &= \frac{1}{k_i} \bigg|\sum_{l=1}^N A_{il} e^{\mathrm{j}\phi_l}\bigg|,
	\end{split}
	\label{eq:KO}
\end{equation}
where $\phi_i$, $\omega_i$ and $k_i$ are the phase, natural frequency and the degree of the $i^{th}$ oscillator respectively, also $\mathrm{j}=\sqrt{-1}$. The parameter $\lambda$ is the overall coupling strength, and the matrix $A=[A_{il}]$ is the underlying graph adjacency. $R_i$ represents the local order parameter that contributes adiabatically to the coupling term and provides the mechanism for explosive synchronization. The values of $\omega_i$ are uniformly distributed over the range $[\omega_0-\frac{\Delta}{2},\omega_0+\frac{\Delta}{2}]$, where $\omega_0$ is the central frequency and $\Delta$ is the width of the frequency range. To quantify the network's coherence, the authors used the averaged global order parameter as
\begin{equation}
	R = \frac{1}{N(t_{max}-t_{trans})}\int_{t_{trans}}^{t_{max}}\Big|\sum_{l=1}^N e^{\mathrm{j}\phi_l(t)}\Big|dt,
	\label{eq:Order}
\end{equation}
$t_{trans}$ and $t_{max}$ being the transient time and the maximal simulation time, respectively.

\begin{figure}[!t]
	\centerline{\includegraphics[scale=0.7500]{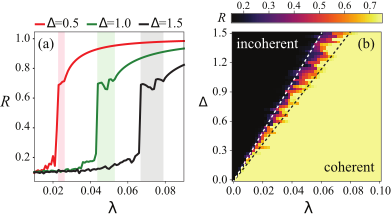}}
	\caption{(a) $R$ as a function of $\lambda$ in the non-locally coupled network of $N=100$ oscillators with $p=0.0$ and $k=10$ for different values of the natural frequency distribution width: $\Delta=0.5$ (red); $\Delta=1.0$ (green); $\Delta=1.5$ (black). Shadings highlight the respective areas of partially coherent chimera-like regimes. (b) Phase diagram in the $(\lambda,\Delta)$ parameter plane with respect to the global order parameter $R$. We refer the reader to the ref. \cite{frolov2020chimera} for further details. }\label{fig4}
\end{figure}

\par For homogeneous frequency distribution (i.e., for $\Delta=0$), the system \eqref{eq:KO} goes through a smooth transition to coherence with weak coupling, however, heterogeneous distribution yields explosive transition to coherence. Besides, a finite-size plateau has been identified where the system undergoes a partially coherent state with the averaged order parameter $R\approx0.7$. The path of transition does not depend on the level of heterogeneity $\Delta$ (cf. fig.\ \ref{fig4}). The impact of the continuous variation of $\Delta$ and $\lambda$ is portrayed in fig.\ \ref{fig4}(b). The region between the dashed white and black lines reflects the existence of chimera-like behavior. Evidently, this interval that supports the chimera-like state improves considerably as $\Delta$ increases. It has further been established that the observed chimera-like state is excited under weakly non-local, small-world, and sparse scale-free coupling and suppressed in globally coupled, strongly rewired, and dense scale-free networks (see \cite{frolov2020chimera} for the detailed mechanism of the evolution of attractive and repulsive mean couplings which is responsible for such chimeric patterns).  
\par Finally, to conclude, we have put forward a brief review to point out to the readers about the recent developments on the field of attractive-repulsive interactions in networks of coupled dynamical systems. We have discussed a few articles of the existing literature, as covering all of them is beyond the scope of this overview. In spite of that, we have presented the dynamical scenarios emerging due to the simultaneity of attraction and repulsion as thorough as possible atleast from the perspective of diversity of the reviewed collective states. We have explained how collective states such as chimera states, solitary states, extreme events, amplitude (or, oscillation) death, anti-phase synchrony, cluster states, travelling waves, different $\pi$-states can appear in ensembles of oscillatory units subject to the coaction of attractive and repulsive interactions. In this context, several relevant challenges lie ahead which will help to bring new insights into this interdisciplinary topic. For instance, further precise strategies are needed to implement for achieving anti-phase synchronization on larger networks and multilayer networks under the influence of such mixed attractive-repulsive coupling to overcome the limit on network size as recognized in the ref.\ \cite{vathakkattil2020limits}. Another important challenge in this direction will be to find out analytically the OD state particularly in large networks depending on mixed interactions between species. A very crucial question from the theoretical perspective is whether it is possible to explore further the existential criterion and properties of Daido's oscillator glass transition \cite{daido1992quasientrainment}. In fact, it will be interesting if another simplified model can be constructed which is capable of revealing such a glass transition with mixed attractive and repulsive interactions. Motivated by analogies to spin glasses as well as to rumor propagation, the microscopic mechanism of the diverse cognitive processes with both positive and negative couplings may provide new insights into some aspects of interacting units. In fact, a careful design on such frustrated networks must be emphasized in order to mitigate the expected extreme events. Along this line of research, a controlling appraoch is proposed in the refs.\ \cite{chowdhury2019extreme,chowdhury2020distance}. But, a generic scheme to restrict such catastrophic events is yet to be found for such coupled systems with mixed interactions. The possibility of forecasting the occurrence of such large events is still an open challenge, and thus the development of new tools is really essential. Besides, controlling chimera or solitary-like weak-chimera states has been one of the most challenging tasks and has not been well-attempted yet, particularly in networked systems experiencing both attraction and repulsion. We hope this short review on attractive-repulsive interactions will open new venues for a better understanding of the underlying mechanism of different emergent states in coupled systems. 
\medskip
\acknowledgments
S.N.C. would  like  to  acknowledge  the  CSIR  (Project No. 09/093(0194)/2020-EMR-I) for financial assistance.



\end{document}